\documentclass[%
 reprint,
 amsmath,amssymb,
 aps,
pra,
]{revtex4-2}

\usepackage{graphicx}
\graphicspath{ {Figures/} }

\usepackage{dcolumn}
\usepackage{bm}
\usepackage{bbold}
\usepackage[export]{adjustbox}

\usepackage{mathtools} 
\DeclarePairedDelimiter\bra{\langle}{\rvert}
\DeclarePairedDelimiter\ket{\lvert}{\rangle}
\DeclarePairedDelimiterX\braket[2]{\langle}{\rangle}{#1 \delimsize\vert #2}

\usepackage{array,colortbl,xcolor}

\usepackage[utf8]{inputenc}
\usepackage[T1]{fontenc}

\usepackage{mathrsfs}
\usepackage{dsfont}

\usepackage{amssymb}
\usepackage{amsmath}
\usepackage{amsfonts}
\usepackage{amsthm}

\usepackage{hyperref}
\hypersetup{pdfpagemode=UseNone}

\newcounter{rem}
\renewcommand{\b}[1]{\mathbf{#1}}

\newcommand{\mc}[1]{\mathcal{#1}}

\def\>{\rangle}
\def\<{\langle}
\newcommand{\proj}[1]{| #1 \rangle\! \langle #1 |}
\newcommand{\idty}{\mathds{1}}
\def\tr{{\rm tr}}
\def\pr{{\rm Pr}}

\def\ii{{\rm i}}

\def\textbf#1{{\bf #1}}

\newcommand{\Nl}{\mathbb{N}}

\newcommand{\Rl}{\mathds{R}}

\usepackage{xcolor}

\usepackage{soul}

\begin{document}


\title{Macro-to-micro quantum mapping and the emergence of nonlinearity}
\author{Pedro Silva Correia}
\email{pedrosc8@cbpf.br}
\affiliation{Centro Brasileiro de Pesquisas F\'isicas, Rio de Janeiro, Rio de Janeiro, CEP 22290-180}
\author{Paola Concha Obando}
\email{pcobando@cbpf.br}
\affiliation{Centro Brasileiro de Pesquisas F\'isicas, Rio de Janeiro, Rio de Janeiro, CEP 22290-180}
\author{Raúl O.~Vallejos}
\email{vallejos@cbpf.br}
\affiliation{Centro Brasileiro de Pesquisas F\'isicas, Rio de Janeiro, Rio de Janeiro, CEP 22290-180}
\author{Fernando de Melo}
\email{fmelo@cbpf.br}
\affiliation{Centro Brasileiro de Pesquisas F\'isicas, Rio de Janeiro, Rio de Janeiro, CEP 22290-180}
\date{\today}
\begin{abstract}
	As a universal theory of physics, quantum mechanics must assign states to every level of description of a system -- from a full microscopic description, all the way up to an effective macroscopic characterization -- and also to  describe the interconnections among them. Assuming that we only have a  coarse-grained access to a physical system, here  we show how to assign to it a microscopic description that abides by all macroscopic constraints. In order to do that, we employ general coarse-graining maps, allowing our approach to be used even when the split between system and environment is not obvious. As a by-product, we show how effective nonlinear dynamics can emerge from the linear quantum evolution, and we readily apply it to a state discrimination task.
\end{abstract}
\maketitle

\section{Introduction}

Quantum mechanics, currently assumed to be  a fundamental theory of physics, was originally formulated to explain the atomic and subatomic scales of nature~\cite{bohr1928}. However, as the atomic hypothesis asserts the  macroscopic world to be composed by a collection of such small constituents, quantum mechanics inherits a universal status: it must be able to explain phenomena at all levels of description. Put differently, quantum mechanics must assign states at both microscopic level, when a complete characterization of the underlying physical system is assumed, and at a macroscopic level of description, which is given by few effective (coarse-grained) degrees of freedom that we have access to.  This universality implies a two-way describing of nature. Firstly, given a macroscopic description of a system,  we must relate it to all possible microscopic states that are compatible with our observations. Secondly, the theory  must show how macroscopic behaviors emerge from microscopic features, even if the latter does not express the behavior of the former. 

The first direction, from macro-to-micro, is commonly covered by quantum statistical physics~\cite{balian2007}. In such approach the main paradigm is that of a ``small'' system of interest interacting with a ``large'' environment, about which we do not have control. Given the values of some physical properties of the system, like its mean internal energy or temperature, statistical physics asserts that the best description of the system is given by the canonical ensemble.
In this open quantum  system paradigm~\cite{caldeira,zurek2006,breuer2002}, as we don't have access to the  environmental degrees of freedom, and we often assume thermal equilibrium and weak interaction, correlations between system and environment can be safely ignored. However, there are situations where the split between accessible and inaccessible degrees of freedom is not so obvious and correlations build up, leading for example to non-Markovian evolutions~\cite{rivas2014, breuercolloquium,ines2017,paula2016,nadja2015exp,pollock2018}.

In the opposite direction, from micro-to-macro, the general idea is to keep only the effective degrees of freedom that we can manipulate. In the open quantum system scenario mentioned above, to get rid of the environment we trace out its degrees of freedom via the partial-trace map. However, when such a split between system and environment is not possible~\cite{cris2017,pedrinho}, the partial-trace has to  be generalized~\cite{alicki2009, cris2017,pedrinho,oleg2020a}.

\begin{figure}
	\centering
	\begin{tabular}{ll}
		& $\mathrm{MACRO}$\hspace*{3cm}$\mathrm{MICRO}$ \\
		(a) & \includegraphics[width=7.5cm,valign=t]{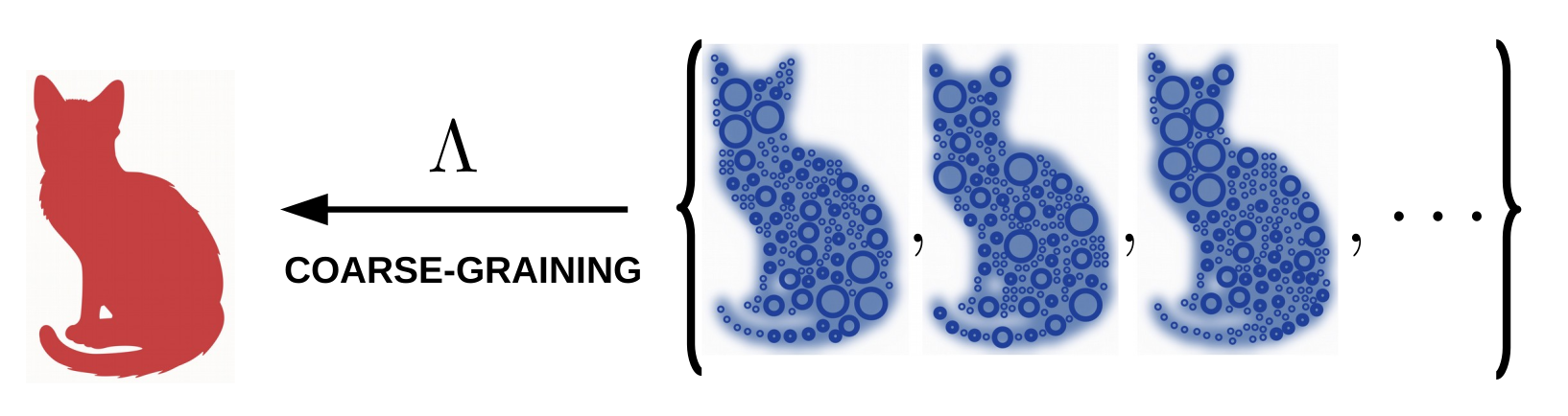} \\\\
		(b) & \includegraphics[width=7.5cm,valign=t]{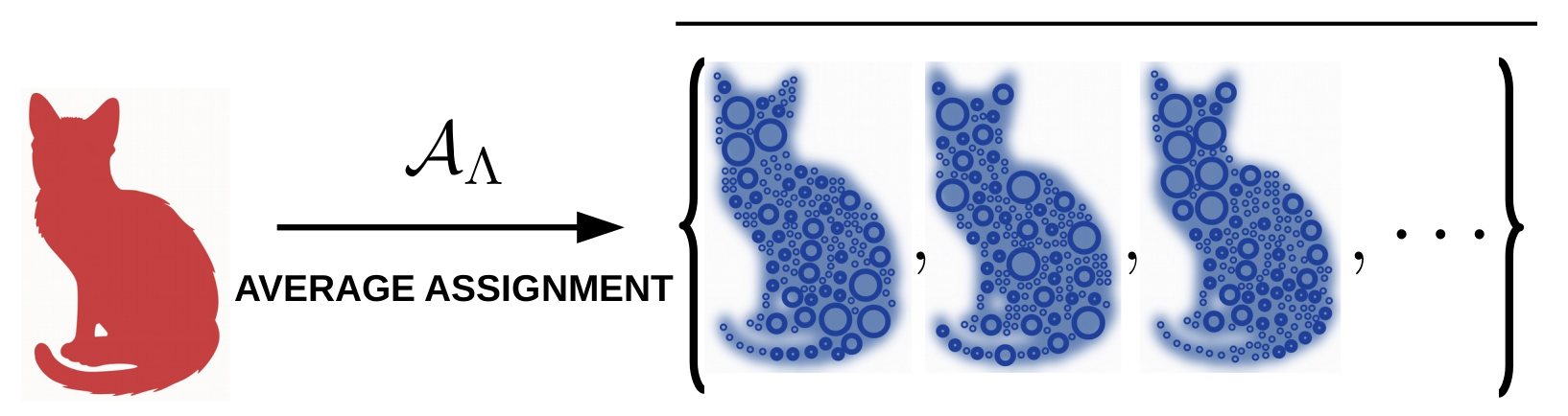} 
	\end{tabular}
	\caption{
		(a) \textbf{Micro-to-macro assignment.}
		In the right side, we pictorially represent in blue the set all microscopic states that are mapped through the coarse-graining operation 
		$\Lambda$ to a unique effective macroscopic state --  represented by the red cat in the left. 
		(b) \textbf{Macro-to-micro assignment}. 
		The map $\mathcal{A}_\Lambda$ assigns to the system an ensemble given by the average over  all microscopic states that comply with the macroscopic observations.}
	\label{fig:cat}
\end{figure}

Here we tap on the theory of coarse-graining maps~\cite{mermin1980, poulin2005, caslavLG, Raeisi2011, Wang2013,Jeong2014, Park2014, cris2017,pedrinho,oleg,cris2019,isadora2020,gabriel2020} in order to consistently describe how quantum mechanics deals with the micro-to-macro and macro-to-micro assignments, even when the distinction between system and environment is not clear cut. Considering an arbitrary set of macroscopic observations, our method gives an ensemble of microscopic states which the underlying physical system could be in. Such an approach is grounded on two basics premises: i) Our  perception of nature invariably arises through measurement processes -- whether considering the everyday perception of our surrounding environment or a sophisticated experimental setup, physical systems are fundamentally perceived and characterized in terms of measurement results of physical observables. ii) Our macroscopic perception of the world is inherently coarse-grained, with  ``classical'' features emerging due to an effective description of quantum systems. 
Based on these premises, our method assigns to a set of macroscopic observations a microscopic description which is the ensemble-average of all microscopic states that are compatible with that observations (see Fig.~\ref{fig:cat}).

As a by-product of our two-way generalization of the open quantum system paradigm, we show how nonlinear dynamics may effectively emerge from the linear quantum evolution. Hitherto, effective quantum nonlinear dynamics were obtained via system specific approximations~\cite{gross1961,pitaevskii1961}, by assignments that intrinsically employed nonlinear functions~\cite{karen2004}, or by continuous measurement of the system~\cite{scott2001}, and as such they lacked a general framework within quantum mechanics.  Given the ubiquity of nonlinear processes within the macro world~\cite{steven1994nonlinear}, such a general microscopic understanding is as desirable as necessary.

\section{Coarse-graining maps} 

Supported by the quantum channel formalism~\cite{MichaelGuide,nielsenchuang}, a coarse-graining operation is characterized  by a completely positive trace preserving (CPTP) map that preserves only the accessible effective degrees of freedom. Let $\mc{H}_a$ be a Hilbert space with  $\dim(\mc{H}_a)=a\in \Nl$, and $\mc{L}(\mc{H}_a)$ the linear operators acting on $\mc{H}_a$.  A coarse-graining map $\Lambda:\mc{L}(\mc{H}_D)\rightarrow\mc{L}(\mc{H}_d)$ connects systems descriptions with different number of degrees of freedom in such a way that $d < D$. Given a state $\psi\in\mc{L}(\mc{H}_D)$ describing a microscopic system with a large number of degrees of freedom, its related coarse-graining description, where only a few effective degrees of freedom are taken into account, is  $\rho=\Lambda[\psi]\in\mc{L}(\mc{H}_d)$. 

A usual coarse-graining situation addressed in the literature is found in the open quantum system scenario~\cite{breuer2002}. In this case, the coarse-graining map responsible for reducing the description to only the system of interest is the partial trace $\tr_E:\mc{L}(\mc{H}_S\otimes\mc{H}_E)\rightarrow\mc{L}(\mc{H}_S)$, where $\mc{H}_S$ is the subspace related to the system and $\mc{H}_E$ the subspace of the environment. Despite successful, the open quantum system approach is limited in its ability to describe other physical scenarios. 

A coarse-graining approach, besides including the partial trace,  can also model isolated quantum systems submitted to measurements with imperfect resolution~\cite{mermin1980, poulin2005, caslavLG, Raeisi2011, Wang2013,Jeong2014, Park2014, cris2017,pedrinho, isadora2020, gabriel2020}. One of such situations is that of cold atoms in a optical lattice~\cite{gross2017quantum,fig1bloch,fukuhara}. In this type of experiments the spin state is detected by fluorescence imaging technique~\cite{sherson}, where the system is illuminated by a global laser pulse and a microscope detects the light scattered by excited atoms (spin-down states $\ket{1}$), while non-excited atoms (spin-up states $\ket{0}$) do not scatter light. To model, in a simplified manner, the scenario where two neighboring atoms cannot be resolved, e.g., the case of a blurred and saturated detection, the coarse-graining map $\Lambda_\text{BnS}:\mc{L}(\mc{H}_4)\rightarrow\mc{L}(\mc{H}_2)$ was put forward in~\cite{cris2017,pedrinho}.   As suggested by the experimental method of fluorescence measurement, this map takes the description of two atoms and converts it into an effective single atom state. A coarse-graining map modeling the basic features of this experimental situation is given below (details can be found in~\cite{cris2017,pedrinho}):
\small
\begin{equation}
\begin{tabular}{lc|cl}
$\Lambda_\text{BnS}[\ket{00}\bra{00}]
=\ket{0}\bra{0}$ &&& $\Lambda_\text{BnS}[\ket{10}\bra{00}]
=\frac{1}{\sqrt{3}}\ket{1}\bra{0}$ \\

$\Lambda_\text{BnS}[\ket{00}\bra{01}]
=\frac{1}{\sqrt{3}}\ket{0}\bra{1}$ &&& $\Lambda_\text{BnS}[\ket{10}\bra{01}]
=0$ \\

$\Lambda_\text{BnS}[\ket{00}\bra{10}]
=\frac{1}{\sqrt{3}}\ket{0}\bra{1}$ &&& $\Lambda_\text{BnS}[\ket{10}\bra{10}]
=\ket{1}\bra{1}$ \\

$\Lambda_\text{BnS}[\ket{00}\bra{11}]
=\frac{1}{\sqrt{3}}\ket{0}\bra{1}$ &&& $\Lambda_\text{BnS}[\ket{10}\bra{11}]
=0$ \\

$\Lambda_\text{BnS}[\ket{01}\bra{00}]
=\frac{1}{\sqrt{3}}\ket{1}\bra{0}$ &&& $\Lambda_\text{BnS}[\ket{11}\bra{00}]
=\frac{1}{\sqrt{3}}\ket{1}\bra{0}$ \\

$\Lambda_\text{BnS}[\ket{01}\bra{01}]
=\ket{1}\bra{1}$ &&& $\Lambda_\text{BnS}[\ket{11}\bra{01}]
=0$ \\

$\Lambda_\text{BnS}[\ket{01}\bra{10}]
=0$ &&& $\Lambda_\text{BnS}[\ket{11}\bra{10}]
=0$ \\

$\Lambda_\text{BnS}[\ket{01}\bra{11}]
=0$ &&& $\Lambda_\text{BnS}[\ket{11}\bra{11}]
=\ket{1}\bra{1}$
\label{eq:CGblurredmap}
\end{tabular}
\end{equation}
\normalsize

Two points about $\Lambda_\text{BnS}$
are worth stressing: First, it cannot be seen as the partial-trace of either one of the two atoms -- note that both $\ket{01}\!\bra{01}$ and $\ket{10}\!\bra{10}$ are mapped to $\ket{1}\!\bra{1}$. Second, the coherence terms within the excited subspace, $\text{span}(\{\ket{01},\ket{10},\ket{11}\})$, vanish in the coarse-grained description because these vectors cannot be discriminated by the detection process, and thus a relative phase between them plays no role -- see~\footnote{Take for example a microscopic state $\ket{\psi} = (\ket{01}+ e^{\ii \phi} \ket{10})/ \sqrt{2}$. This state, microscopically, does present a coherence between the states $\ket{01}$ and $\ket{10}$. Now, assuming that both $\proj{01}$ and $\proj{10}$ are mapped to $\proj{1}$, if the effective coherence term does not vanish, the result of applying the coarse-graining map on $\proj{\psi}$, i.e., $\Lambda_{\mathrm{BnS}}[\proj{\psi}]$ wouldn’t be a valid quantum state.} for further explanation.

Both coarse-graining maps, $\tr_E$ and $\Lambda_\text{BnS}$, are going to be used throughout this article in order to illustrate the main concepts and differences between the traditional open quantum system approach and the more general coarse-graining one.

\section{Averaging assignment maps} 

Now we aim to address the opposite direction: we want to define a procedure that maps a macroscopic (coarse-grained) description of a system, with $d$ degrees of freedom, to a microscopic one, with $D$ degrees of freedom (with $d < D$). We assume  that the macroscopic description is obtained from the microscopic one via a coarse-graining map, say $\Lambda:\mc{L}(\mc{H}_D)\rightarrow\mc{L}(\mc{H}_d)$.  Then, given a physical system described by a set $\mc{O}=\{o_i\}$ of $N_\mc{O}$ mean values, quantum mechanics assigns coarse-grained observables $O_i \in \mc{L}(\mc{H}_d)$, and microscopic pure quantum states $\psi:= \proj{\psi}\in \mc{L}(\mc{H}_D)$, such that $o_i= \tr[\Lambda[\psi] O_i]$. 

In the scenario pictured above, notice that the microscopic state $\psi$ satisfying the macroscopic constraints $\mathcal{O}$ is in general not unique. We then define the set of all possible microscopic pure quantum states that abide by the macroscopic constraints:
\begin{equation}
	\Omega_{\Lambda}(\mc{O})=\big\{\psi\in\mc{L}(\mathcal{H}_D)\;\big|\;\tr[O_i\Lambda[\psi]\big]=o_i\,,\forall\,1{\leq}\,i\,{\leq}N_\mc{O}\big\}.
	\label{eq:omegaO}
\end{equation}

In an operational perspective, when assembling an effective preparation  with properties $\mc{O}$, which is accessed through a coarse-graining map $\Lambda$,  microscopically we are in fact sampling from the set $\Omega_{\Lambda}(\mc{O})$.  Due to the linearity of the expectation value, this perspective suggests an averaging map $\mc{A}_\Lambda: \mc{O}\rightarrow  \mc{L}(\mc{H}_D)$ that assigns the appropriate description to the microscopic ensemble:
\begin{align}
\mathcal{A}_{\Lambda}[\mc{O}]&\equiv\overline{\Omega_{\Lambda}(\mc{O})}^\psi={\int} d{\mu_\psi}\pr_\Lambda(\psi|\mc{O})\,\psi,
\label{eq:avgO}
\end{align}
where  $d{\mu_\psi}$ is the uniform Haar measure over pure states, and  $\pr_\Lambda(\psi|\mc{O})$ is the probability density of having the microscopic state $\psi$ given the macroscopic constraints imposed by $\mc{O}$ and the coarse-graining map $\Lambda$. Note that $\pr_\Lambda(\psi|\mc{O})=0$ for any $\psi \not\in  \Omega_{\Lambda}(\mc{O})$. 

In the particular case where the set $\mc{O}$ is big enough as to allow for the full state reconstruction of $\rho$ in $\mc{L}(\mc{H}_d)$, i.e. $\mc{O}$ is tomographically complete, then we can see $\mc{A}_\Lambda$ as a map between states, $\mc{A}_\Lambda: \mc{L}(\mc{H}_d) \rightarrow  \mc{L}(\mc{H}_D)$. This assignment map, in general is not completely positive nor linear, but these characteristics pose no problems as we discuss in what follows.

\subsection{Averaging assignment: open quantum system}
 
We start by applying our formalism to the traditional open quantum scenario where $\Lambda=\tr_E$. We assume that the state of the system is completely known and described by $\rho_S\in\mc{L}(\mc{H}_S)$. No further constraints are assumed. In this case, the set of pure states of the whole system and environment, $\Omega_{\tr_E}(\rho_S)=\{\psi\in\mc{L}(\mc{H}_S\otimes\mathcal{H}_E)|\tr_E[\psi]=\rho_S\}$, is formed by the purifications of $\rho_S$. As no further constraints are imposed,  each purification appears with the same probability in $\Omega_{\tr_E}(\rho_S)$. Evaluating (\ref{eq:avgO}) for the partial-trace case, we have:
\begin{equation}
\mathcal{A}_{\tr_E}[\rho_S]=\rho_S\otimes\dfrac{\mathds{1}}{d_E},
\label{eq:avtrace1}
\end{equation}
with $d_E=\dim(\mc{H}_E)$. This calculation is shown in Appendix~\ref{ap:apptrace}. For this choice of coarse-graining map, the averaging assignment map ${A}_{\tr_E}$ is  linear in $\rho_S$ and completely positive. 

\subsection{Averaging assignment: blurred and saturated detector} 

Now we turn our attention to the case of a blurred and saturated detection as described by the coarse-graining map $\Lambda_\text{BnS}$ defined in~\eqref{eq:CGblurredmap}. As before, given $\rho\in\mc{L}(\mc{H}_2)$ we want to take the average over the states belonging to the set $\Omega_{\Lambda_\text{BnS}}(\rho)=\{\psi\in\mc{L}(\mc{H}_4)| \Lambda_\text{BnS}[\psi]=\rho \}$. 
Writing the elements of $\rho$ in the computational basis as $\rho_{ij}=\<i|\rho\ket{j}$, for $i,j\in\{0,1\}$, the averaging assignment map $\mc{A}_{\Lambda_\text{BnS}}[\rho]$ gives:
\begin{equation}
\begin{pmatrix}
\rho_{00} & \dfrac{\rho_{01}}{\sqrt{3}} & \dfrac{\rho_{01}}{\sqrt{3}} & \dfrac{\rho_{01}}{\sqrt{3}} \\
\dfrac{\rho_{01}^\ast}{\sqrt{3}} & \dfrac{\rho_{11}}{3} & \dfrac{|\rho_{01}|^2}{2\rho_{00}}-\dfrac{\rho_{11}}{6} & \dfrac{|\rho_{01}|^2}{2\rho_{00}}-\dfrac{\rho_{11}}{6}\\
\dfrac{\rho_{01}^\ast}{\sqrt{3}} & \dfrac{|\rho_{01}|^2}{2\rho_{00}}-\dfrac{\rho_{11}}{6} & \dfrac{\rho_{11}}{3} & \dfrac{|\rho_{01}|^2}{2\rho_{00}}-\dfrac{\rho_{11}}{6} \\
\dfrac{\rho_{01}^\ast}{\sqrt{3}} & \dfrac{|\rho_{01}|^2}{2\rho_{00}}-\dfrac{\rho_{11}}{6} & \dfrac{|\rho_{01}|^2}{2\rho_{00}}-\dfrac{\rho_{11}}{6} & \dfrac{\rho_{11}}{3}
\end{pmatrix}.
\label{eq:avblurred}
\end{equation}
The calculation is detailed in Appendix~\ref{ap:apblurred}.

Observe that differently from the average state related to the partial trace case  (\ref{eq:avtrace1}), for the blurred and saturated detector the assigned average state is nonlinear on $\rho$. The nonlinear terms are related to the coherences within the excited subspace $\text{span}(\{\ket{01}, \ket{10},\ket{11}\})$,  which the detector cannot resolve. All these nonlinear terms vanish after the coarse-graining map is applied. This nonlinearity, nevertheless, can be expressed in a dynamical process. 

\section{Effective state dynamics} 

Now, supported by the averaging assignment procedure, in this section we will establish an operational procedure that characterizes how stochastic effective dynamics arise from deterministic unitary quantum dynamics. The goal is similar to the one analyzed in~\cite{cris2017}. Here, however, we assume  access only to the coarse-grained description of the system, to a model of the microscopic dynamics, and to the coarse-graining map describing our ability  to construe the microscopic world. More explicitly, given the initial effective state $\rho(0)\in\mc{L}(\mc{H}_d)$, the microscopic unitary evolution map $\mc{U}_t:\mc{L}(\mc{H}_D)\rightarrow \mc{L}(\mc{H}_D)$, and the coarse-graining map $\Lambda:\mc{L}(\mc{H}_D)\rightarrow \mc{L}(\mc{H}_d)$, we want to construct a family of effective channels $\Gamma_t:\mc{L}(\mc{H}_d)\rightarrow \mc{L}(\mc{H}_d)$ such that for each time $t\in \Rl_+$ the evolved effective state is given by $\rho(t) = \Gamma_t[\rho(0)]$.

\begin{figure}
	\centering
	\begin{tabular}{lr}
		(a) & \includegraphics[width=7.3cm,valign=t]{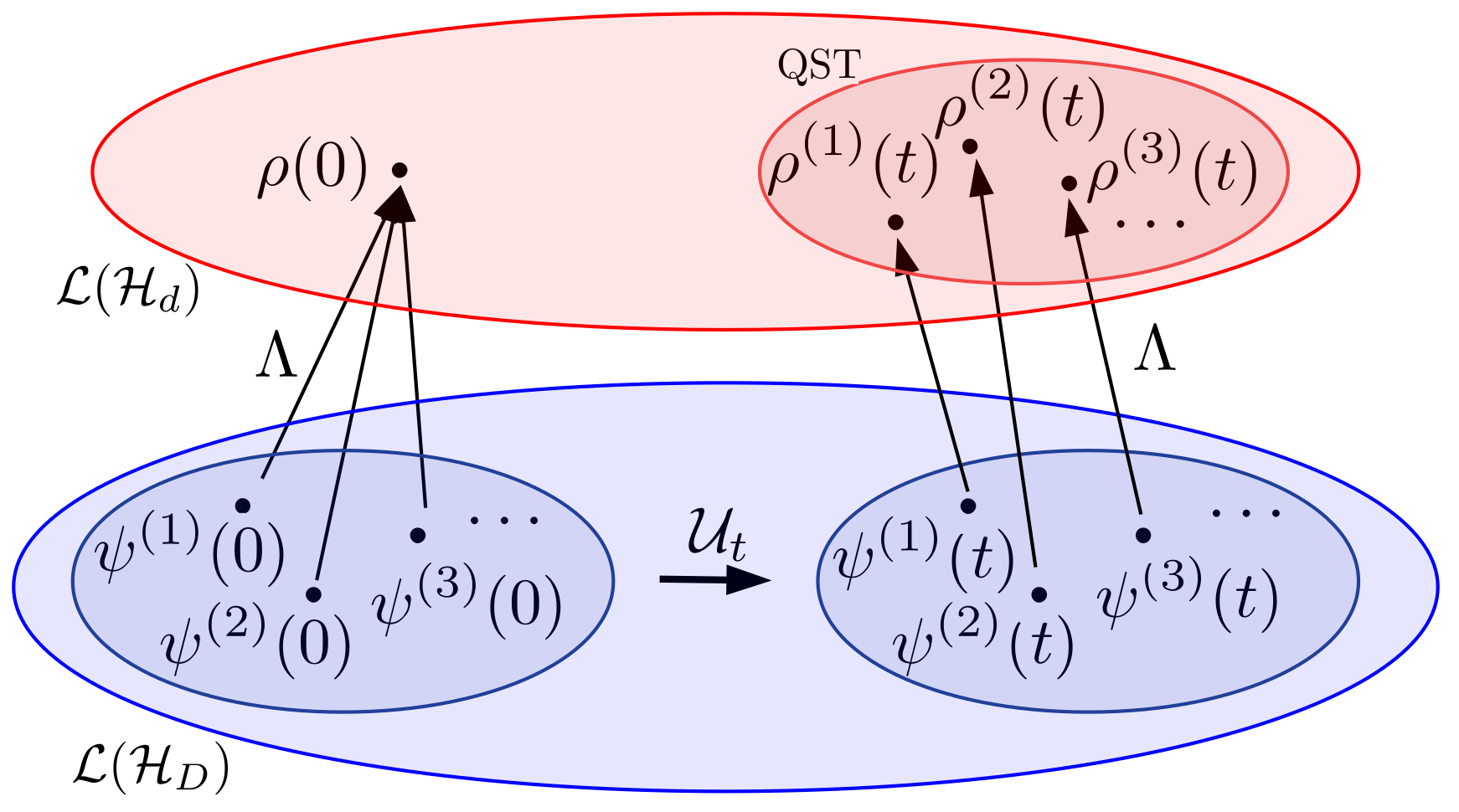} \\\\
		(b) & \includegraphics[width=7.3cm,valign=t]{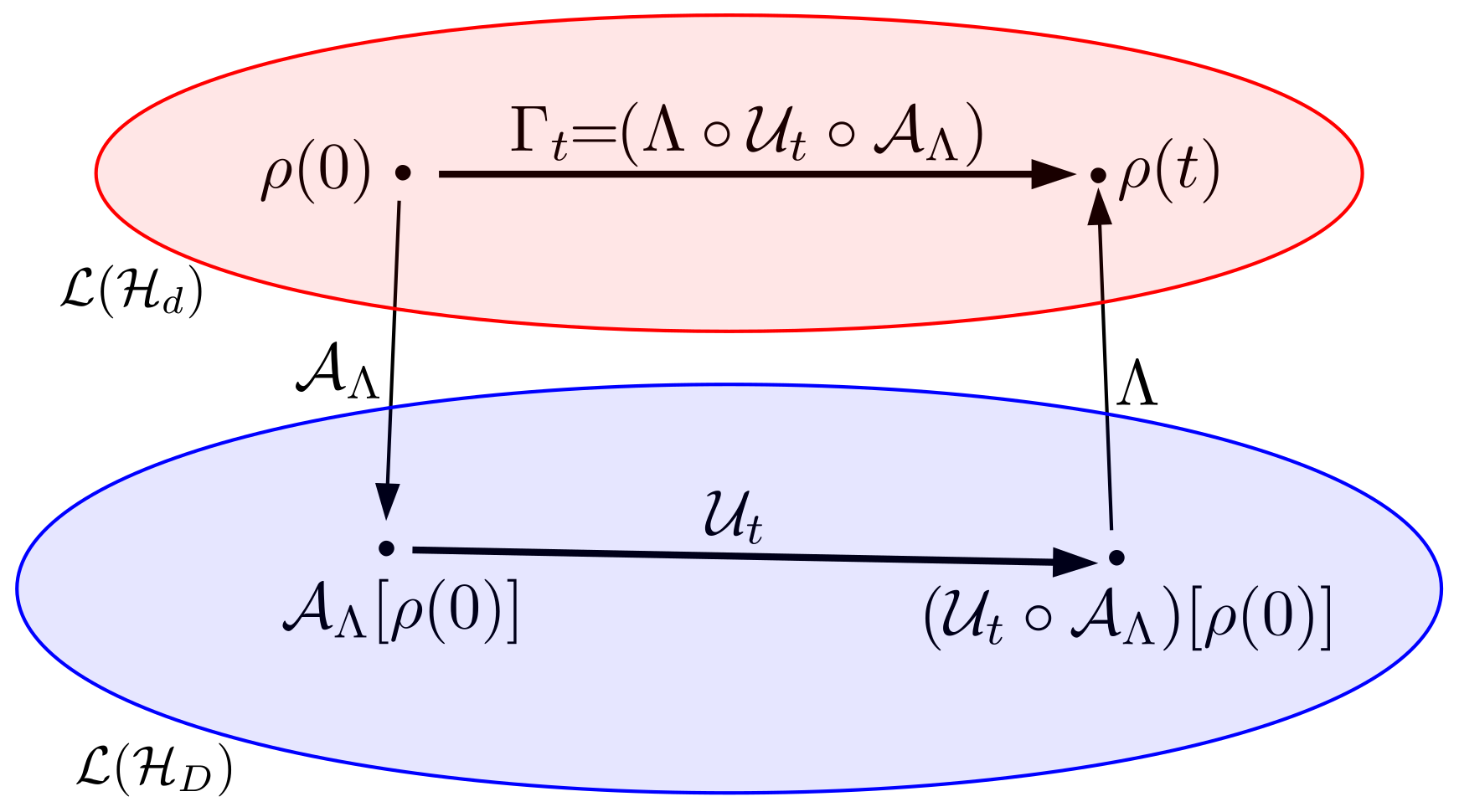} 
	\end{tabular}
	\caption{(a) \textbf{Single run evolution.} In the $i$-th run, the preparation of $\rho(0)$ implies the random preparation of a microscopic state $\psi^{(i)}(0) \in \Omega_{\Lambda}(\rho(0))$,
which then evolves according to the unitary map $\mc{U}_t$, and through the coarse-graining map $\Lambda$ finally gives the effective state $\rho^{(i)}(t)= (\Lambda\circ\mc{U}_t)[\psi^{(i)}(0)]$.  In each run a possibly different effective state is created.
(b)\textbf{ Effective evolution.} The scheme in (a), together with the linearity of quantum mechanics, suggests an effective dynamics given by $\Gamma_t=(\Lambda\circ\mathcal{U}_t\circ\mathcal{A}_{\Lambda})$.}
	\label{fig:CGdynamics}
\end{figure}

Once again we appeal to the operational mindset in order to obtain $\Gamma_t$. To prepare the initial effective  state $\rho(0)\in\mathcal{L}(\mathcal{H}_d)$ means, in each run, to prepare a microscopic state from the set $\Omega_{\Lambda}(\rho(0))=\{\psi\in\mathcal{L}(\mathcal{H}_D)|\Lambda[\psi]=\rho(0)\}$. Let $\psi^{(i)}(0)\in \Omega_{\Lambda}(\rho(0))$ be the microstate selected, with probability $\pr_\Lambda(\psi^{(i)}(0)|\rho(0))$, in the $i$-th run. Microscopically, this state evolves through the unitary map $\mc{U}_t$, and then to obtain its effective description we apply the coarse-graining map $\Lambda$. All this leads, in the $i$-th run, to $\rho^{(i)}(t)= (\Lambda\circ\mc{U}_t)[\psi^{(i)}(0)]$ -- see Fig.~\ref{fig:CGdynamics}~(a). If we are to determine the final effective state, for instance via a quantum state tomography (QST), this procedure has to be performed many times, and an averaging naturally appears:
\begin{align}
\rho(t)&=\int d\mu_{\psi(0)}\pr_\Lambda(\psi(0)|\rho(0))\; (\Lambda\circ\mathcal{U}_t)[\psi(0)]\nonumber\\
&=(\Lambda\circ\mathcal{U}_t)\left[\int d\mu_{\psi(0)}\pr_\Lambda(\psi(0)|\rho(0))\psi(0)\right]. 
\label{eq:averagerhot2}
\end{align}
In the second line above we used the linearity of both the unitary evolution and coarse-graining channel. Note that the integral in \eqref{eq:averagerhot2}  is the very definition of the averaging assignment map acting on $\rho(0)$. Combining all the three steps, the effective dynamical map $\Gamma_t:\mc{L}(\mc{H}_d)\rightarrow\mc{L}(\mc{H}_d)$ is obtained as:
\begin{align}
\Gamma_t=(\Lambda\circ\mathcal{U}_t\circ\mathcal{A}_{\Lambda}).
\label{eq:gammat}
\end{align}
The construction of the effective  coarse-grained dynamical model is schematically represented in Fig.~\ref{fig:CGdynamics}~(b). 

\subsection{Effective state dynamics: open quantum system} 
In the partial trace case, the description of the whole system and environment fulfilling the constraint is given by the averaging assignment map, $\mc{A}_{\tr_E}[\rho(0)]=\rho(0)\otimes\frac{\mathbb{1}}{d_E}$. From (\ref{eq:gammat}), in this open quantum system scenario, the effective evolved state $\rho_t$ is obtained as follows:
\begin{equation}
\rho(t)=(\tr_{E}\circ\mathcal{U}_t)[\rho(0)\otimes\frac{\mathds{1}}{d_E}].
\label{eq:gammatrace}
\end{equation}
Note that the effective dynamics in this case is linear on $\rho(0)$, since both the unitary evolution $\mc{U}_t$ and the coarse-graining map $\tr_E$ are linear operations. Moreover, as $\mc{U}_t$ and $\tr_E$ are completely positive, so it is $\Gamma_t$.

We analyze now a simple scenario, a local unitary evolution. With $\mathcal{U}_t=\mathcal{U}_t^S\otimes\mathcal{U}_t^E$, the unitary evolution: $
\rho(t)=(\tr_{E}\circ\mathcal{U}_t^S\otimes\mathcal{U}_t^E)[\rho(0)\otimes\frac{\mathds{1}}{d_E}]=\mathcal{U}_t^S[\rho(0)]$ is recovered, so the effective evolution $\Gamma_t=\mathcal{U}_t^S$, more than linear, it is unitary. 

\subsection{Effective state dynamics: blurred and saturated detector.} 
As previously discussed, the average state in the detector case,  Eq.\eqref{eq:avblurred}, has a nonlinear dependence on the effective state. This  may lead to a nonlinear effective dynamics if we allow the composed system to evolve before the application of the coarse-graining map. 
For the nonlinearity in the average state to imply a nonlinear effective dynamics, an appropriate unitary evolution must be chosen so that the nonlinear term  in (\ref{eq:avblurred}) shows up in elements other than the single excitation subspace coherences, as those vanish by the action of $\Lambda_\text{BnS}$.

As an example, consider a unitary evolution generated by the local Hamiltonian $H=\hbar\omega \idty\otimes\sigma_y$, i.e., the unitary evolution of the averaged assigned state is given by  $\mc{U}_t[\mc{A}_{\Lambda_\text{BnS}}[\rho(0)]]=U_t\mc{A}_{\Lambda_\text{BnS}}[\rho(0)]U_t^\dagger$ with $U_t=\exp[-i\omega t(\idty\otimes\sigma_y)]$. Despite of being a local evolution,  with respect to the microscopic system, the effective dynamics is nonlinear in this case. Take for instance the initial state $\rho(0)$ with components $\rho_{00}(0)=\rho_{11}(0)=1/2$ and $\rho_{01}(0)\in \Rl_+$, to simplify the analysis. Its evolution leads to an effective state whose probability to be found in $\ket{0}$ at $\omega t= \pi/3$ is given by 
$(1-2\rho_{01}(0))^2/16$. 
It clearly shows a nonlinear dependence on the initial state. See Appendix~\ref{ap:BNSdynamics} for details.

Lastly, note that this effective channel is not a usual input-output  black-box from the quantum channel theory~\cite{MichaelGuide}. As such the definition of complete positivity does not directly apply here. Nevertheless, from the operational interpretation of this effective channel, it is clear that even when we extend the microscopic system by adding auxiliary systems, the effective evolution will always produce a valid quantum state. This comes from the complete positivity of the unitary and coarse-graining maps, allied with the sampling from the set of states that abide by the macroscopic constraints.

\section{Application: Effective state discrimination}
 
Non-linear processes have far-reaching applications, ranging from optics to the description of biological systems~\cite{steven1994nonlinear}.  As we now show, the  nonlinear dynamics shown above can  be useful in the task of discriminating between two effective states. 

One  foundational result in quantum information theory is the so-called Helstrom bound~\cite{helstrom1969}. It states that if  a source has the same chance of preparing one out of two states, say $\rho$ and $\chi$ in  $\mc{L}(\mc{H})$, then the maximum probability of a measurement to correctly identify which state was produced is $\left(1 +\mc{D}(\rho, \chi)\right)/2$,
where $\mc{D}(\rho,\chi)=\tr|\rho-\chi|/2$ is the trace distance between the states. The larger the distance between the two states, the bigger is the probability of distinguishing between them. Another central result of quantum information theory is that the distance between  two states does not increase by the action of a linear CPTP map $\Phi: \mc{L}(\mc{H})\rightarrow \mc{L}(\mc{H}^\prime)$, i.e., $\mc{D}(\Phi[\rho],\Phi[\chi])\le \mc{D}(\rho,\chi)$. As such, the probability of distinguishing between two states cannot increase by further linear processing of the system.

\begin{figure}
	\centering
	\includegraphics[width=\linewidth]{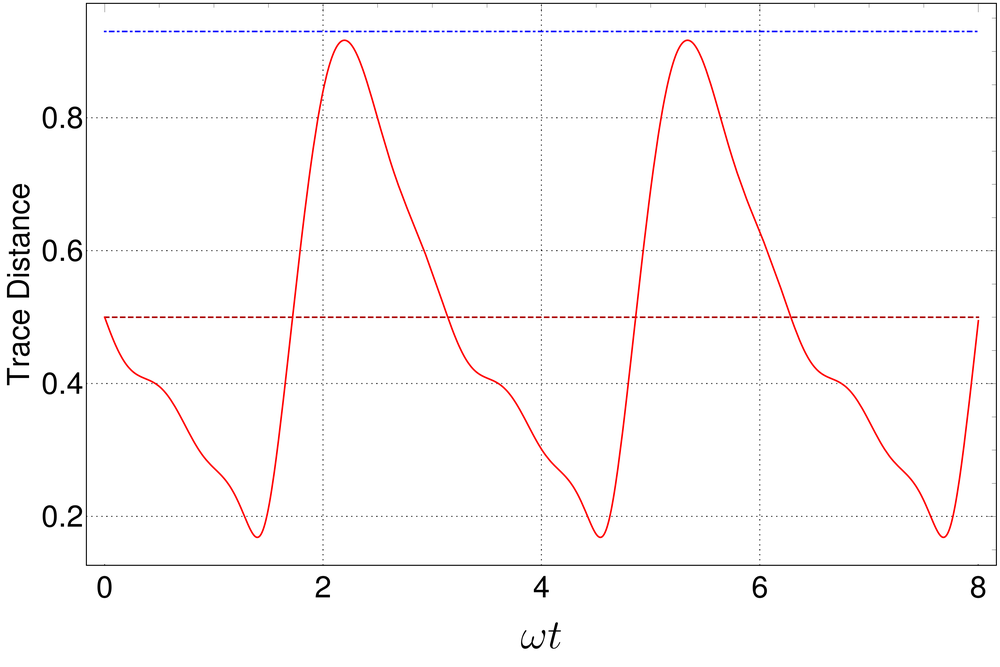}
	\caption{\textbf{Effective distance evolution.} The distance between two effective system's descriptions can increase due to a nonlinear coarse-grained dynamics. The red solid line describes the distance evolution $\mc{D}(\rho(t),\chi(t))$; the red dashed line represents the initial distance $\mc{D}(\rho(0),\chi(0))$; while the blue dot-dashed line shows the distance between the underlying assignments $\mc{D}(\mc{A}_{\Lambda_\text{BnS}}[\rho(0)],\mc{A}_{\Lambda_\text{BnS}}[\chi(0)])$.}
	\label{fig:distance}
\end{figure}

Now assume that we want to distinguish between two effective states, and we have access to a nonlinear dynamics as shown above. In this case the probability of discriminating between two effective states can in fact increase. For concreteness, assume that $\rho(0) = \idty/2$ and $\chi(0) =\proj{\chi}$ with $\ket{\chi}= \sqrt{0.8}\ket{0}+\sqrt{0.2}\ket{1}$. Moreover, let the coarse-graining channel  be the  $\Lambda_\text{BnS}$ introduced above, and the microscopic dynamics  be governed by the Hamiltonian $H=\hbar\omega(\idty\otimes\sigma_y+\sigma_y\otimes\idty)$. In this case, the effective channel $\Gamma_t$ is nonlinear, and as shown in Fig.~\ref{fig:distance} the distance among the two effective states can be bigger than its initial value, allowing therefore for a better effective state discrimination.

It must be stressed that the increase in state discrimination is only possible in the coarse-grained description. It is simple to see that 
$\mc{D}(\Gamma_t[\rho(0)],\Gamma_t[\chi(0)])\le \mc{D}(\mc{A}_{\Lambda_\text{BnS}}[\rho(0)],\mc{A}_{\Lambda_\text{BnS}}[\chi(0)])$,
and thus the best effective discrimination is never better than the best microscopic discrimination. If it were possible to increase the success of discriminating between two states by throwing away some information,  besides  astonishing, it would also violate the non-signaling principle~\cite{gisin1990, MichaelGuide}.

\section{Conclusions} 

We generalized the usual statistical mechanics scenario, which presumes a clear split between system and environment, to an extended definition of subsystems. We formalized and generalized, within quantum theory, the microscopic ensemble assignment fixed a set of macroscopic constraints. Our approach allows for a decoherence-like reasoning to be applied even to closed systems. Not only the locality of observables matter, but also  how coarse-grained is our level of description. Even a theoretically isolated cat is rarely described by the state of its subatomic particles. 

From this approach it is clear how stochastic nonlinear effective dynamics may emerge from the deterministic linear quantum evolution. Again the level of description, and thus the  ability to prepare a macroscopic system, is the key. Fixing a coarse-grained preparation is usually not sufficient to determine the underlying state. As we showed, the best suited description of the micro state is possibly nonlinear on the macro state.  This effective nonlinearity can be expressed in a dynamical process, but it introduces no conflict with the physical tenets of complete positivity and no-signaling.

We see the present framework as a first step towards a more general statistical physics, which may have impact on foundational aspects -- like the quantum-to-classical transition --, but also in more applied topics --  such as in quantum communication protocols that exploit the discrimination between effective states.

\section*{Acknowledgments} 

We gladly acknowledge fruitful discussions with Christiano Duarte. This work is supported by the Brazilian funding agencies CNPq and CAPES, and it is part of the Brazilian National Institute for Quantum Information.

\onecolumngrid

\appendix

\section{Averaging assignment: open quantum system}
\label{ap:apptrace}

Here we explicitly evaluate the  average state related to the partial trace case. The method to be presented here will serve as inspiration for the calculation of the average state related to the blurred and saturated detector.

Given the partial trace map $\tr_E:\mc{L}(\mc{H}_S\otimes\mc{H}_E)\rightarrow\mc{L}(\mc{H}_S)$ and a coarse-grained description $\rho\in\mc{L}(\mc{H}_S)$, the average state $\mathcal{A}_{\tr_{E}}[\rho]$ is given by
\begin{equation}
	\mc{A}_{\tr_E}[\rho] = \int d\mu_{\psi} \pr_{\tr_E}(\psi|\rho)\psi;
\end{equation}
where $d\mu_{\psi}$ is the uniform Haar measure in $\mc{H}_S\otimes\mc{H}_E$, and the conditional probability $\pr_{\tr_E}(\psi|\rho)$ is non-null only in  $\Omega_{\tr_E}(\rho)=\{\ket{\psi}\in\mc{H}_S\otimes\mc{H}_E\,|\,\tr_{E}[\proj{\psi}]=\rho\}$.  The elements of $\Omega_{\tr_E}(\rho)$ are thus the purifications of $\rho$ in the extended space $\mc{H}_S\otimes\mc{H}_E$.

In order to abide by the coarse-graining constraint, the conditional probability distribution $\pr_{\tr_E}(\psi|\rho)$ must be proportional to $\delta(\tr_E[\psi]-\rho)$. Such a probability is invariant by unitary transformations in the ``environment'' part, that is:
\begin{equation}
	\pr_{\tr_E}(\psi|\rho)=\pr_{\tr_E}(\mathds{1}{\otimes}U\psi\mathds{1}{\otimes}U^\dagger|\rho), \; \forall U\in\mc{L}(\mc{H}_E).
\end{equation}
As the purifications of $\rho$ are connected by local unitary transformations  acting  in $\mc{H}_E$, this invariance implies that all the elements in $\Omega_{\tr_E}(\rho)$, given no further constraints, are equally likely.

Since the Haar measure $d\mu_\psi$ is also invariant by unitary transformations,  the average state for the partial trace case can be equivalently written by changing the variables $\ket{\psi}\rightarrow\mathds{1}{\otimes}U\ket{\psi}$ as:
\begin{align}
	\mathcal{A}_{\tr_E}[\rho]={\int}d\mu_{\psi} \pr_{\tr_E}(\psi|\rho)\,\mathds{1}{\otimes}U\psi\mathds{1}{\otimes}U^\dagger.
	\label{eq:avestatetrace1}
\end{align} 
Given the choice of unitary $U$ in the equation above plays no role, we can average over all such unitary transformations, to obtain:
\begin{align}
	\mathcal{A}_{\tr_E}[\rho]={\int}d\mu_{\psi} \pr_{\tr_E}(\psi|\rho)\,\overline{\mathds{1}{\otimes}U\psi\mathds{1}{\otimes}U^\dagger}^{\mu_U}.
	\label{eq:avestatetrace2}
\end{align}

As we have established an equal probability for all states in $\Omega_{\tr_E}(\rho)$, this average is performed using the Haar measure on the environment part, and its explicit evaluation is a standard result in quantum information~\cite{MichaelGuide}:
\begin{equation}
	\overline{\mathds{1}{\otimes}U\psi\mathds{1}{\otimes}U^\dagger}^{\mu_U}=\tr_{E}[\psi]\otimes\dfrac{\mathds{1}}{d_E}=\rho\otimes\dfrac{\mathds{1}}{d_E}.
	\label{eq:avtrace}
\end{equation}
Note that the above result is independent of $\psi$, depending only on the coarse-grained density matrix $\rho$. The integral in (\ref{eq:avestatetrace2}) is then now easily calculated:
\begin{align}
	\mathcal{A}_{\tr_E}[\rho]&=\rho\otimes\dfrac{\mathds{1}}{d_E}\overbrace{{\int}d\mu_\psi\,\pr_{\tr_E}(\psi|\rho)}^{=1} \nonumber \\
	&=\rho\otimes\dfrac{\mathds{1}}{d_E}
	\label{eq:avestatetrace3}.
\end{align}
Therefore, in the partial trace case, the average state is just the tensor product between the coarse-grained state $\rho$ and the identity in subspace $\mc{H}_E$. 

\section{Averaging assignment: blurred and saturated detector}
\label{ap:apblurred}

Here we calculate the average state related to the blurred and saturated detector, as induced by the coarse-graining map  $\Lambda_\text{BnS}:\mc{L}(\mc{H}_4)\rightarrow\mc{L}(\mc{H}_2)$. Following the same steps as in the above calculation, the average assignment for the present case is given by:
\begin{equation}
	\mc{A}_{\Lambda_\text{BnS}}[\rho] = \int d\mu_{\psi} \pr_{\Lambda_\text{BnS}}(\psi|\rho)\psi;
\end{equation}
where $d\mu_{\psi}$ is the uniform Haar measure in $\mc{H}_4$, and the conditional probability distribution $\pr_{\Lambda_\text{BnS}}(\psi|\rho)$ is non-null only in  $\Omega_{{\Lambda_\text{BnS}}}(\rho)=\{\ket{\psi}\in\mc{H}_4\,|\,{\Lambda_\text{BnS}}[\proj{\psi}]=\rho\}$.

The coarse-graining constraints imply that $\pr_{\Lambda_\text{BnS}}(\psi|\rho)\propto \delta(\Lambda_\text{BnS}[\proj{\psi}]-\rho)$. Here, however, the symmetry obeyed by this conditional probability distribution is not so immediately spotted.  
In order to make it apparent, we write $\ket{\psi}$ in the computational basis in $\mc{H}_4$ as $\ket{\psi}=\sum_{i,j=0}^{1}c_{ij}\ket{ij}$, where $c_{ij}\in\mathbb{C}$, and $\sum_{ij}|c_{ij}|^2=1$. Consequently, the average assignment can be written as:
\begin{align}
	\mathcal{A}_{\Lambda_\text{BnS}}[\rho]={\int}dc_{00}dc_{01}dc_{10}dc_{11}\pr_{\Lambda_\text{BnS}}(c_{00},c_{01},c_{10},c_{11}|\rho)\,\psi(c_{00},c_{01},c_{10},c_{11})\,\delta(|c_{00}|^2+|c_{01}|^2+|c_{10}|^2+|c_{11}|^2-1).
	\label{eq:avstate2}
\end{align}
Now note that the action of $\Lambda_\text{BnS}$ on $\psi$ is the following one:
\begin{equation}
	\Lambda_\text{BnS}[\psi]=\begin{pmatrix}
		|c_{00}|^2 & 	\frac{1}{\sqrt{3}}c_{00}[c_{01}^*+c_{10}^*+c_{11}^*]\\
		\frac{1}{\sqrt{3}}c_{00}^*[c_{01}+c_{10}+c_{11}] & |c_{01}|^2+|c_{10}|^2+|c_{11}|^2
	\end{pmatrix}.
	\label{eq:cij}
\end{equation}
Therefore, in terms of the coefficients $c_{ij}$, the proportionality $\pr_{\Lambda_\text{BnS}}(\psi|\rho)\propto \delta(\Lambda_\text{BnS}[\proj{\psi}]-\rho)$ can be rewritten as:
\begin{align}
	\pr_{\Lambda_\text{BnS}}(c_{00},c_{01},c_{10},c_{11}|\rho)\propto&\delta\big(|c_{00}|^2-\rho_{00}\big)
	\label{eq:c1}\times \\ 
	&\times\delta\big(|c_{01}|^2+|c_{10}|^2+|c_{11}|^2-\rho_{11}\big)\times
	\label{eq:c2} \\
	&\times\delta\Big(\frac{c_{00}}{\sqrt{3}}[c_{01}^\ast+c_{10}^\ast+c_{11}^\ast]-\rho_{01}\Big),
	\label{eq:c3}
\end{align}
where $\rho_{ij}=\<i|\rho|j\>$ are the components of $\rho$ in the computational basis in $\mc{H}_2$. Note that the normalization restriction $\delta(|c_{00}|^2+|c_{01}|^2+|c_{10}|^2+|c_{11}|^2-1)$ is already implied by the normalization of $\rho$ and the constraints in \eqref{eq:c1} and \eqref{eq:c2}.

The coefficients $c_{ij}$ are complex numbers, and can be thus be written as $c_{ij}=a_{ij}+\ii\,b_{ij}$ with $a_{ij},b_{ij}\in\mathbb{R}$. In this sense, in order to calculate the integral in \eqref{eq:avstate2} it is convenient to rewrite $\ket{\psi}$ as: 
\begin{equation}
	\ket{\psi}=YV,
	\label{eq:vecAB}
\end{equation}
with $Y$ and $V$ respectively defined as: 
\begin{equation}
	Y\equiv\begin{pmatrix}
		1&0&0&0&i&0&0&0 \\
		0&1&0&0&0&i&0&0 \\
		0&0&1&0&0&0&i&0 \\
		0&0&0&1&0&0&0&i
	\end{pmatrix},
	\quad
	V\equiv\begin{pmatrix}
		a_{00}\\a_{01}\\a_{10}\\a_{11}\\b_{00}\\b_{01}\\b_{10}\\b_{11}
	\end{pmatrix}.
\end{equation}
The density matrix representation  is then equivalently written as $\ket{\psi}\bra{\psi}=YVV^TY^\dagger$, with the average state (\ref{eq:avstate2}) expressed now as:
\begin{align}
	\mathcal{A}_{\Lambda_\text{BnS}}[\rho]={\int}dV\; \pr_{\Lambda_\text{BnS}}(V|\rho)\,YVV^TY^\dagger\,\delta( V^T V-1),
	\label{eq:avstate3}
\end{align}
where $dV=\prod_{i,j=0}^{1}da_{ij}db_{ij}$ and the probability $\pr_{\Lambda_\text{BnS}}(V|\rho)\equiv\pr_{\Lambda_\text{BnS}}(a_{00},\dots,a_{11},b_{00},\dots,b_{11}|\rho)$. Without loss of  generality we can ignore a global phase and  consider $c_{00} $ real, such that $a_{00}=c_{00}$ and $b_{00}=0$. With these considerations, $\pr_{\Lambda_\text{BnS}}(V|\rho)$ can be rewritten as the following product of the delta functions:
\begin{align}
	\pr_{\Lambda_\text{BnS}}(V|\rho)\propto\;
	&\delta\big(a_{00}^2-\rho_{00}\big)\times\delta\big(b_{00}\big)\times
	\label{eq:ab1} \\
	&\delta\big(a_{01}^2+a_{10}^2+a_{11}^2+b_{01}^2+b_{10}^2+b_{11}^2-\rho_{11}\big)\times
	\label{eq:absphere} \\
	&\delta\Big(\frac{a_{00}}{\sqrt{3}}(a_{01}+a_{10}+a_{11})-\mathfrak{Re}[\rho_{01}]\Big)\times
	\label{eq:abplane1} \\
	&\delta\Big(-\frac{a_{00}}{\sqrt{3}}(b_{01}+b_{10}+b_{11})-\mathfrak{Im}[\rho_{01}]\Big).
	\label{eq:abplane2}
\end{align}
The first two delta functions in the expression above already fix $a_{00}=\sqrt{\rho_{00}}$, and $b_{00}=0$. The second line, \eqref{eq:absphere}, imposes a spherical symmetry for the real coefficients in the excited subspace, as it is equivalent to a sphere of radius $\sqrt{\rho_{11}}$ in such a subspace. This symmetry suggests the conditional probability  $\pr_{\Lambda_\text{BnS}}(V|\rho)$ to be invariant over orthogonal transformations on the excited subspace.

The orthogonal transformations that leave $\pr_{\Lambda_\text{BnS}}(V|\rho)$ invariant are, however, further restricted by the constraints in Eqs.~\eqref{eq:abplane1} and~\eqref{eq:abplane2}. The allowed transformations are those that maintain the hyper-planes $a_{01}+a_{10}+a_{11}=\sqrt{3}\mathfrak{Re}[\rho_{01}]/a_{00}$ and $b_{01}+b_{10}+b_{11}=-\sqrt{3}\mathfrak{Im}[\rho_{01}]/a_{00}$ invariant. Such transformations are rotations in the corresponding subspaces along vectors normal to the hyper-planes.

We thus established that 
\begin{equation}
	\pr_{\Lambda_\text{BnS}}(V|\rho)=\pr_{\Lambda_\text{BnS}}(O(\theta,\phi) V|\rho).
	\label{eq:pv}
\end{equation}
for orthogonal transformations of the form 
\begin{equation}
	O(\theta,\phi)=\mathbb{1}{\oplus}R_a(\theta){\oplus}\mathbb{1}{\oplus}R_b(\phi)=\begin{pmatrix}
		1 & \begin{array}{ccc}0 & 0 & 0\end{array}& 0 &\begin{array}{ccc}0 & 0 & 0\end{array} \\
		\begin{array}{c}0\\0\\0\end{array} & R_a(\theta) & \begin{array}{c}0\\0\\0\end{array} & \begin{array}{ccc}0 & 0 & 0\\0 & 0 & 0\\0 & 0 & 0\end{array} \\
		0 & \begin{array}{ccc}0 & 0 & 0\end{array}& 1 &\begin{array}{ccc}0 & 0 & 0\end{array}\\
		\begin{array}{c}0\\0\\0\end{array} & \begin{array}{ccc}0 & 0 & 0\\0 & 0 & 0\\0 & 0 & 0\end{array} & \begin{array}{c}0\\0\\0\end{array} & R_b(\phi)
	\end{pmatrix},
	\label{eq:orthogonal}
\end{equation}
where $R_a(\theta)$ is a rotation in the ``$a$'' excited subspace by an angle $\theta\in [0,2\pi[$ along the axis $a=(1,1,1)$, and similarly, $R_b(\phi)$ is a rotation in the ``$b$'' excited subspace by an angle $\phi\in [0,2\pi[$ along the axis $b=(1,1,1)$.

\medskip

Now we can proceed as for the partial trace case. Using the invariance property \eqref{eq:pv} in the average assigned description \eqref{eq:avstate3} we get:
\begin{align}
	\mathcal{A}_{\Lambda_\text{BnS}}[\rho]=&{\int}d(O(\theta,\phi)V)\; \pr_{\Lambda_\text{BnS}}(O(\theta,\phi)V|\rho)\,YO(\theta,\phi)VV^TO^T(\theta,\phi)Y^\dagger \delta( V^T O^T(\theta,\phi) O(\theta,\phi) V-1)\\
	&= {\int}dV\; \pr_{\Lambda_\text{BnS}}(V|\rho)\,YO(\theta,\phi)VV^TO^T(\theta,\phi)Y^\dagger \delta( V^T V-1),
\end{align}
where we used that $d(O(\theta,\phi)V)=dV$ as $O(\theta,\phi)$ is an orthogonal transformation.  As the above equation is true for any choice of $\theta$ and $\phi$, we can uniformly average over these parameters to get: 
\begin{align}
	\mathcal{A}_{\Lambda_\text{BnS}}[\rho]= {\int}dV\; \pr_{\Lambda_\text{BnS}}(V|\rho)\,Y\overline{O(\theta,\phi)VV^TO^T(\theta,\phi)}^{\mu_O}Y^\dagger\,\delta( V^T V-1),
	\label{eq:avestate4}
\end{align}
where $\mu_O$ is the uniform measure over the orthogonal transformations $O(\theta, \phi)$. Explicitly, this averaging can be written as:
\begin{align}
	Y\overline{O(\theta,\phi)VV^TO^T(\theta,\phi)}^{\mu_O}Y^\dagger=Y\bigg(\dfrac{1}{(2\pi)^2}{\int}_{0}^{2\pi}d{\theta}{\int}_{0}^{2\pi}d{\phi}\,O(\theta,\phi)VV^TO^T(\theta,\phi)\bigg)Y^\dagger.
	\label{eq:averegestate2}
\end{align}
Although tedious, the integral can be exactly calculated, and leads to the following matrix:
\begin{equation}
	Y\overline{O(\theta,\phi)VV^TO^T(\theta,\phi)}^{\mu_O}Y^\dagger=\begin{pmatrix}
		\bigcirc & \triangle & \triangle & \triangle \\
		\triangle^\ast & \Diamond & \square & \square\\
		\triangle^\ast & \square & \Diamond & \square \\
		\triangle^\ast & \square & \square & \Diamond \\
	\end{pmatrix},
\end{equation}
with the coefficients $\bigcirc,\Diamond,\triangle$ and $\square$ dependent of $a_{ij}$ and $b_{ij}$ as follows:
\begin{align}
	\bigcirc=&\,a_{00}^2 , \nonumber\\
	\Diamond=&\,\dfrac{1}{3}(a_{01}^2+a_{10}^2+a_{11}^2+b_{01}^2+b_{10}^2+b_{11}^2), \nonumber\\
	\triangle=&\,\dfrac{1}{3}a_{00}(a_{01}+a_{10} + a_{11}-i\,b_{01} - i\,b_{10}-i\,b_{11}), \nonumber\\
	\square=&\,\dfrac{1}{3}(a_{01}a_{10}+a_{01}a_{11}+a_{10}a_{11}+b_{01}b_{10}+b_{01}b_{11}+b_{10}b_{11}).
\end{align}
Employing the constraints in Eqs.\eqref{eq:ab1} - \eqref{eq:abplane2}, these coefficients can be rewritten as
\begin{align}
	\bigcirc&= \rho_{00},& \Diamond&=\dfrac{\rho_{11}}{3}, &\triangle&=\dfrac{\rho_{01}}{\sqrt{3}},&\square&=\frac{3|\triangle|^2}{\bigcirc}-\dfrac{\Diamond}{2}.
\end{align}
With these results, we finally get:
\begin{equation}
	Y\overline{O(\theta,\phi)VV^TO^T(\theta,\phi)}^{\mu_O}Y^\dagger=\begin{pmatrix}
		\rho_{00} & \dfrac{\rho_{01}}{\sqrt{3}} & \dfrac{\rho_{01}}{\sqrt{3}} & \dfrac{\rho_{01}}{\sqrt{3}} \\
		\dfrac{\rho_{01}^\ast}{\sqrt{3}} & \dfrac{\rho_{11}}{3} & \dfrac{|\rho_{01}|^2}{2\rho_{00}}-\dfrac{\rho_{11}}{6} & \dfrac{|\rho_{01}|^2}{2\rho_{00}}-\dfrac{\rho_{11}}{6}\\
		\dfrac{\rho_{01}^\ast}{\sqrt{3}} & \dfrac{|\rho_{01}|^2}{2\rho_{00}}-\dfrac{\rho_{11}}{6} & \dfrac{\rho_{11}}{3} & \dfrac{|\rho_{01}|^2}{2\rho_{00}}-\dfrac{\rho_{11}}{6} \\
		\dfrac{\rho_{01}^\ast}{\sqrt{3}} & \dfrac{|\rho_{01}|^2}{2\rho_{00}}-\dfrac{\rho_{11}}{6} & \dfrac{|\rho_{01}|^2}{2\rho_{00}}-\dfrac{\rho_{11}}{6} & \dfrac{\rho_{11}}{3}\\
	\end{pmatrix}.
	\label{eq:avstate5}
\end{equation}
Note that the above matrix is independent of $V$, i.e., it is independent of $\ket{\psi}$, depending only on the elements of the effective given state $\rho$. As such, the average assignment can be obtained as:

\begin{align}
	\mathcal{A}_{\Lambda_\text{BnS}}[\rho]=  Y\overline{O(\theta,\phi)VV^TO^T(\theta,\phi)}^{\mu_O}Y^\dagger\; \overbrace{{\int}dV\; \pr_{\Lambda_\text{BnS}}(V|\rho) \delta( V^T V-1)}^{=1}=Y\overline{O(\theta,\phi)VV^TO^T(\theta,\phi)}^{\mu_O}Y^\dagger.
	\label{eq:averegestate}
\end{align}
Thus the average state related to the blurred and saturated coarse-graining is given by (\ref{eq:avstate5}).

\section{Effective state dynamics: Blurred and saturated detector}
\label{ap:BNSdynamics}

Here we detail the calculations of the example presented in the main text, related to the effective evolved state in the blurred and saturated case. 

We start by considering the initial coarse-grained state $\rho(0)\in\mathcal{L}(\mathcal{H}_2)$:
\begin{align}
	\rho(0)=\begin{pmatrix}
		\dfrac{1}{2} & \rho_{01}(0) \\
		\rho_{01}(0) & \dfrac{1}{2}
	\end{pmatrix},
\end{align} 
with $\rho_{01}(0)\in \Rl_+$, to simplify the analysis. 

Given that $\rho(t)=(\Lambda_{\mathrm{BnS}}\circ\mathcal{U}_t\circ\mathcal{A}_{\Lambda_{\mathrm{BnS}}})[\rho(0)]$, the first step is the average state $\mathcal{A}_{\Lambda_{\mathrm{BnS}}}[\rho(0)]$, which can be easily constructed following the general form given by Eq.$(5)$ in the main text.  

The next step is the microscopic evolution. We consider the unitary evolution generated by the local Hamiltonian $H=\hbar\omega\idty\otimes\sigma_y$, i.e., the unitary evolution of the averaged assigned state is given by  $\mc{U}_t[\mc{A}_{\Lambda_\text{BnS}}[\rho(0)]]=U_t\mc{A}_{\Lambda_\text{BnS}}[\rho(0)]U_t^\dagger$ with $U_t=\exp[-\ii\omega t(\idty\otimes\sigma_y)]$. 

Finally, we apply the coarse-graining map $\Lambda_{\mathrm{BnS}}$ given by $(1)$ in the main text. Under these circumstances, the evolved coarse-grained effective description at $\omega t=\pi/3$ is given by:  
\begin{align}
	\rho(\pi/3\omega)=\dfrac{1}{16}\begin{pmatrix}
		(1-2\rho_{01}(0))^2 & \dfrac{(1-2\rho_{01}(0))^2(1+2\rho_{01}(0))}{2} \\
		\dfrac{(1-2\rho_{01}(0))^2(1+2\rho_{01}(0))}{2} & 1 - (1-2\rho_{01}(0))^2
	\end{pmatrix}
\end{align} 
Clearly, it shows a nonlinear dependence on the initial state.

\twocolumngrid

%
\end{document}